# Electron Spin Dynamics in a Quantum Dot due to Hyperfine Coupling with Nuclei in the Dot – An Exact Analysis

A. K. Rajagopal, Naval Research Laboratory, Washington, DC 20375

## Abstract

The time-dependent Schrodinger equation of a many particle spin system consisting of an electron in a quantum dot interacting with the spins of the nuclei (N) in the dot due to hyperfine interaction is solved formally exactly for a given arbitrary initial state. The electron spin dynamics is then expressed in terms of the reduced density matrix of the composite system by computing the marginal density matrix of the electron. This is accomplished by classifying the states of the system by the total spin of the coupled electron and nuclear system that commutes with the system Hamiltonian. These states are used in enumerating and finding the exact solution of the time-dependent Schrodinger equation. In each sector of the total spin, the problem reduces to solving a linear simultaneous set of equations that are solved by matrix inversion. Such solutions enable one to make a reliable approximate scheme for purposes of numerical estimation of the various physical quantities. A mathematical and physical discussion of this procedure with the density matrix approach to this problem is given here. To elucidate the procedure and its advantages, special cases of N=3 and 4 are given in some detail in an Appendix.

## I. Introduction

We consider a conduction electron (spin 1/2 for simplicity) in a quantum dot (QD) coupled to N nuclear spins (here for simplicity of presentation, only a single species of nuclei are considered also with spin 1/2) in the QD by the Fermi contact interaction. All the spins are expressed in units of 1/2 and so the constants appearing in the foregoing are appropriately scaled. Typically N is of the order of $10^5$. An external magnetic field B is assumed to be present along a z-direction that orients the spins of both the electron and the nuclei by the Zeeman interaction. The Hamiltonian $H = H_e + H_n + H_{en}$ describes this composite system of electron, nuclei, and their interaction correspondingly where

$$H_e = e\, \mathbf{s}_z \otimes I_n; \quad H_n = I_e \otimes \sum_{k=1}^{N} e_N\, I_{k,z}; \quad H_{en} = \frac{1}{2} \sum_{k=1}^{N} A_k\, \vec{I}_k \cdot \vec{s} \quad (1)$$

Here $e = m_{eB}\, g\, B$ is the electron Zeeman energy with $m_{eB}$ is the electron Bohr magneton with g the g-factor (2, for free electron but is different in semiconductors etc.) $I_n$ is the unit operator in the Hilbert space of nuclei. $e_N = m_{nB}\, g_n\, B$ is the nuclear counterpart of the Zeeman energy. Typically it is smaller in magnitude compared with the electron Zeeman energy but since there are N nuclei, the total contribution may be comparable in magnitude and may even be of opposite sign if the nuclei are oriented antiparallel to the electron spin direction. $I_e$ is the unit operator in the Hilbert space of the electron. The last term represents the hyperfine interaction of the nuclei with the electron with the strength of the interaction $A_k$, (in general positive) depends on the fixed position of the nucleus and the electron density at that location. It is important to note that in a QD, the nuclei are located at fixed lattice points of the underlying crystal structure of the QD. This means that the hyperfine constants $A_k$'s are not random numbers. Here the spin operators of electron and the nuclei are the Pauli operators. Typically, they are $s_x = \begin{pmatrix} 0 & 1 \\ 1 & 0 \end{pmatrix}$, $s_y = \begin{pmatrix} 0 & -i \\ i & 0 \end{pmatrix}$, $s_z = \begin{pmatrix} 1 & 0 \\ 0 & -1 \end{pmatrix}$. We also employ the identity $\vec{I}_k \cdot \vec{s} = I_{kz} s_z + \frac{1}{2}(I_{k+} s_- + I_{k-} s_+)$. Here $a_\pm = a_x \pm i a_y$.

The purpose of this paper is to present formally complete solutions of this Hamiltonian from which certain approximation schemes for numerical estimations of the physical effects entailed in this problem may be deduced in a systematic and reliable way. In the second section we outline the general features of the method of solution. In this process, we introduce some important notations required in the enumeration of the solution. Following this, in the third section, we set up the relevant equations in partitioned matrix form. The solution requires inversion of partitioned matrices and these are elucidated with one example in the fourth section that has been examined in the literature in different ways. In the final section some approximation schemes are outlined that emerge from this analysis. Also a discussion of the density matrix method in contrast to the one presented here will be given in this section. In Appendix A, we illustrate our technique for two illustrative cases where the number of nuclei is 3 and 4. For a comprehensive review of the subject one may consult [1, 2].

## II. General Method of Solution

The time evolution of the total system of electron and the nuclei is determined by the Schrodinger equation (units with Planck constant, $\hbar = h/2\pi$ set equal to unity) $i\partial_t |\Psi(t)\rangle = H |\Psi(t)\rangle$, given some arbitrary initial (t=0) configuration of the electron and nuclear spins. The observation that the total spin of the coupled electron and spin system given by $M_z = s_z + \sum_{k=1}^{N} I_{k,z}$ commutes with the Hamiltonian given by eq.(1) will be used in enumerating and finding the exact solution of this Schrodinger equation. From this solution, we can obtain physical quantities of interest such as the time evolution of various components of electron spin for a given initial condition. We first observe that the solutions for different values of $M_z$ are mutually orthogonal so that we can enumerate solutions for each value of $M_z$. In each sector of $M_z$ one has thus a finite matrix problem. Since all the nuclei and the electron are assumed to be spin 1/2, we may enumerate the states of the system with various possible values of $M_z$. The total number of states is $2^{N+1}$. The enumeration proceeds as follows: consider the electron spin to be along the z-axis, i.e., +1. Then there is one state where all the nuclei are pointing up, for which $M_z = N+1$. With the electron spin +1, there are N possible configurations of one nuclear spin antiparallel to the (N-1) nuclei, for which $M_z = N-1$ However there is one more state with this value of $M_z$ where the electron spin is antiparallel to all the nuclei pointing along the z direction. Thus we may proceed to enumerate all the states as follows:

| | Number of States $N(M_z)$ | $M_z$ - value | States |
|---|---|---|---|
| | 1 | N+1 | $\lvert \uparrow; \Uparrow \cdots \Uparrow \rangle$ |
| | $\binom{N}{0} = 1$ | N-1 | $\lvert \downarrow; \Uparrow \cdots \Uparrow \rangle$ |
| | $\binom{N}{1} = N$ | N-1 | $\{\lvert \uparrow; \Uparrow \cdots \Downarrow_k \cdots \Uparrow \rangle\}$ k=1,2,…,N |
| total | $\binom{N}{0} + \binom{N}{1}$ | N-1 | Two sets together |
| | $\binom{N}{1} = N$ | N-3 | $\{\lvert \downarrow; \Uparrow \cdots \Downarrow_k \cdots \Uparrow \rangle\}$ k=1,2,…,N |
| | $\binom{N}{2} = \frac{N(N-1)}{2}$ | N-3 | $\{\lvert \uparrow; \Uparrow \cdots \Downarrow_{k1} \cdots \Downarrow_{k2} \cdots \Uparrow \rangle\}$ k1,k2=1,2,…,N($k1 \neq k2$) |
| total | $\binom{N}{1} + \binom{N}{2}$ | N-3 | Two sets together |

More generally

$$\binom{N}{m} \qquad \text{N-2m-1} \qquad \{\lvert\downarrow;\Uparrow\cdots\Downarrow_{k1}\cdots\Downarrow_{km}\cdots\Uparrow\rangle\}$$

$$\binom{N}{m+1} \qquad \text{N-2m-1} \qquad \{\lvert\uparrow;\Uparrow\cdots\Downarrow_{k1}\cdots\Downarrow_{k(m+1)}\cdots\Uparrow\rangle\}$$

total $\binom{N}{m}+\binom{N}{m+1}$ \qquad N-2m-1 \qquad Two sets together

And finally

$$\binom{N}{N}=1 \qquad -(N+1) \qquad \lvert\downarrow;\Downarrow\cdots\Downarrow\rangle$$

In the above the combinatorial factor $\binom{N}{m}=\dfrac{N!}{(N-m)!m!}$ is the number of ways of choosing the m nuclear down states among the total number N of nuclei.
Total number of states obtained

$$=\binom{N}{0}+\binom{N}{0}+\binom{N}{1}+\binom{N}{1}+\binom{N}{2}+\ldots+\binom{N}{N-1}+\binom{N}{N}+\binom{N}{N}=2^{N+1}$$

as it should be.

We may thus state that for a given $M_z(m) = N - 2m - 1$, there are $N(M_z(m)) = \binom{N}{m}+\binom{N}{m+1}$ states and so we may express the corresponding state

$$\lvert\Psi_{M_z(m)}(t)\rangle = \sum_{\binom{N}{m}} Y_{\{m\}}(t)\lvert\downarrow;\{m\}\rangle + \sum_{\binom{N}{m+1}} X_{\{m+1\}}(t)\lvert\uparrow;\{m+1\}\rangle \qquad (2)$$

In eq.(2), the coefficients are the respective amplitudes of the corresponding states. Here we have introduced the notation $|\{m\}\rangle \equiv \left|\{\Downarrow\}_m \{\Uparrow\}_{N-m}\right\rangle$, with m=0,1,…,N. The sums go over the indicated combinatorial arrangements of the nuclear spins. The most general state vector is then a linear combination of all these $2^{N+1}$ states.

$$|\Psi(t)\rangle = \sum_{M_z(m)} C(M_z(m)) |\Psi_{M_z(m)}(t)\rangle$$ that is normalized to one by requiring

$$\sum_{M_z(m)} |C(M_z(m))|^2 \left\{ \sum_{\binom{N}{m+1}} |X_{\{m+1\}}(t)|^2 + \sum_{\binom{N}{m}} |Y_{\{m\}}(t)|^2 \right\} = 1 \quad (3)$$

The density matrix associated with this time-dependent pure state of the combined electron-nuclear system is given by $r(t) = |\Psi(t)\rangle\langle\Psi(t)|$. By taking the trace over all the nuclei we obtain the time-dependent marginal density matrix of the electron spin which may be expressed succinctly as

$$r_e(t) = \frac{1}{2}(I_e + \vec{s}(t)\cdot\vec{s}) \quad (4)$$

Here the various spin components are expressed in terms of the amplitudes defined above, which are found by solving the Schrodinger equation, and are given by the following expressions:

$$s_z(t) = \sum_{M_z(m)} |C(M_z(m))|^2 \left\{ \sum_{\binom{N}{m+1}} |X_{\{m+1\}}(t)|^2 - \sum_{\binom{N}{m}} |Y_{\{m\}}(t)|^2 \right\} \quad (5a)$$

$$s_+(t) = \sum_{M_z(m)} \sum_{M_z(m-1)} C^*(M_z(m)) C(M_z(m-1)) \sum_{\binom{N}{m}} X^*_{\{m\}}(t) Y_{\{m\}}(t) \quad (5b)$$

$$s_-(t) = (s_+(t))^* \quad (5c)$$

Here we point out that the z-component of electron spin arises from the solutions appropriate to a given $M_z(m)$ but the transverse component involves two adjacent values of $M_z(m)$ because this involves the overlap of up and down electron spin for which the appropriate nuclear spin configurations appear. The spin dynamics is thus determined in terms of the solutions for any given initial state.

## III. Equations for the amplitudes associated with a fixed $M_z(m)$

We now turn our attention to solving the Schrodinger equation corresponding to the Hamiltonian of eq.(1) for a fixed $M_z(m) = N - 2m - 1$. This consists of two sets of states given in the Table above:

One set consisting of $\binom{N}{m}$ states of the form $\{\lvert\downarrow;\{m\}\rangle \equiv \lvert\downarrow;\{\Downarrow\}_m\rangle\}$ with electron spin down and a corresponding set of $\binom{N}{m+1}$ states of the form $\{\lvert\uparrow;\{m+1\}\rangle \equiv \lvert\uparrow;\{\Downarrow\}_{m+1}\rangle\}$ with electron spin up. From eq.(2), the equations obeyed by the amplitudes are found to be

$$i\dot{Y}_{\{m\}} = \langle\downarrow,\{m\}|H|\Psi(M_z(m))\rangle$$

$$= \sum_{\binom{N}{m'}}\langle\downarrow;\{m\}|H|\downarrow;\{m'\}\rangle Y_{\{m'\}} + \sum_{\binom{N}{m'+1}}\langle\downarrow;\{m\}|H|\downarrow;\{m'\}\rangle X_{\{m'+1\}}$$

(6a)

The last term in eq.(6a) is just $\sum_{\{m'+1\}}\sum_{k=1}^{N}\frac{A_k}{4}\langle\{m\}|I_{+k}|\{m'+1\}\rangle X_{\{m'+1\}}$.

$$i\dot{X}_{\{m+1\}} = \langle\uparrow;\{m+1\}|H|\Psi(M_z(m))\rangle$$

$$= \sum_{\binom{N}{m'}}\langle\uparrow;\{m+1\}|H|\downarrow;\{m'\}\rangle Y_{\{m'\}} + \sum_{\binom{N}{m'+1}}\langle\uparrow;\{m+1\}|H|\uparrow;\{m'+1\}\rangle X_{\{m'+1\}}$$

(6b)

The first term in eq.(6b) is just $\sum_{\{m'\}}\sum_{k=1}^{N}\frac{A_k}{4}\langle\{m+1\}|I_{-k}|\{m'\}\rangle Y_{\{m'\}}$.

We first set up the action of the Hamiltonian given in eq.(1) on a typical state where for convenience we choose to label one of the $\{m\}$ nuclear spins as $\lvert\downarrow;(\Downarrow_1\Downarrow_2\cdots\Downarrow_m)(\Uparrow_{m+1}\cdots\Uparrow_N)\rangle$; there are $\binom{N}{m}$ such sets for a given m. The relevant part of the hyperfine interaction Hamiltonian on this state is $\sum_{k=1}^{N}\left(\frac{A_k}{4}\right)I_{k-}S_{+}$ which flips the electron spin to up state and flips each of the
(N – m) $\Uparrow$ nuclear spins labeled (m+1) to N to a $\Downarrow$ state. To see this clearly we give here the result:

$$H\left|\downarrow;(\Downarrow_1\Downarrow_2\cdots\Downarrow_m)(\Uparrow_{m+1}\cdots\Uparrow_N)\right\rangle =$$

$$-\left(\Omega-\sum_{l=1}^{m}A_l\right)\left|\downarrow;(\Downarrow_1\Downarrow_2\cdots\Downarrow_m)(\Uparrow_{m+1}\cdots\Uparrow_N)\right\rangle \quad (6a')$$

$$+\sum_{l=m+1}^{N}\left(\frac{A_l}{4}\right)\left|\uparrow;(\Downarrow_1\Downarrow_2\cdots\Downarrow_m\Downarrow_l)(\Uparrow_{m+1}\cdots\Uparrow_N)'\right\rangle$$

where we have introduced $\Omega = e + A$, $A = \sum_{k=1}^{N}\frac{A_k}{2}$. The prime indicates that the nuclear spin $\Uparrow_l$ is not in the set because it has been flipped. This suggests a notation

$$\left|\uparrow;\{m+1\}\right\rangle_{\Uparrow_l} \equiv l-\text{th nuclear spin} \Uparrow \text{ is flipped} \quad (7)$$

so that this state now contains (m+1) nuclear $\Downarrow$ spins. Then eq.(6) is written in the form

$$H\left|\downarrow;\{m\}\right\rangle = -\left(\Omega-\sum_{l=1}^{m}A_l\right)\left|\downarrow;\{m\}\right\rangle + \sum_{l=m+1}^{N}\left(\frac{A_l}{4}\right)\left|\uparrow;\{m+1\}\right\rangle_{\Uparrow_l} \quad (6a'')$$

There are $\binom{N}{m}$ such equations for each choice of m – nuclear spins.

Similarly

$$H\left|\uparrow;(\Downarrow_1\Downarrow_2\cdots\Downarrow_{m+1})(\Uparrow_{m+2}\cdots\Uparrow_N)\right\rangle =$$

$$\left(\Omega-\sum_{l=1}^{m+1}A_l\right)\left|\uparrow;(\Downarrow_1\Downarrow_2\cdots\Downarrow_{m+1})(\Uparrow_{m+2}\cdots\Uparrow_N)\right\rangle \quad (6b')$$

$$+\sum_{l=1}^{m+1}\left(\frac{A_l}{4}\right)\left|\downarrow;(\Downarrow_1\Downarrow_2\cdots\Downarrow_m)'(\Uparrow_l\Uparrow_{m+1}\cdots\Uparrow_N)\right\rangle$$

The prime here indicates that the nuclear spin $\Downarrow_l$ is not in the set because it has been flipped. This suggests a notation

$$\left|\downarrow;\{m\}\right\rangle_{\Downarrow_l} \equiv l-\text{th nuclear spin}\Downarrow \text{ is flipped} \quad (8)$$

so that this state now contains (m) nuclear $\Downarrow$ spins. Equation (9) is then written in a compact way

$$H|\uparrow;\{m+1\}\rangle = \left(\Omega - \sum_{l=1}^{m+1} A_l\right)|\uparrow;\{m+1\}\rangle + \sum_{l=1}^{m+1}\left(\frac{A_l}{4}\right)|\downarrow;\{m\}\rangle_{\Downarrow_l} \quad (6b')$$

There are $\binom{N}{m+1}$ similar equations for each of the choices of (m+1) nuclear spins. The entire sets of these equations in conjunction with eq.(2) determine the amplitudes for the m – value chosen above for which the state has
$M_z(m)$ = N - 2m - 1. For the above configuration of m nuclear spins, we have the following set of equations:

$$i\dot{Y}_{\{m\}}(t) = -B_{\{m\}}Y_{\{m\}}(t) + \sum_{l=m+1}^{N}\left(\frac{A_l}{4}\right)X_{\{m+1\},l}(t) \quad (9)$$

$$i\dot{X}_{\{m+1\}}(t) = B_{\{m+1\}}X_{\{m+1\}}(t) + \sum_{l=1}^{m}\left(\frac{A_l}{4}\right)Y_{\{m\}/l}(t) \quad (10)$$

We introduced some compact notations here in anticipation of generalizing the equations to arbitrary choice of m nuclei. Thus, $B_{\{m\}} = e + A - \sum_{l=1}^{m} A_l$; $X_{\{m+1\},l}(t)$ means that the l – th $\Uparrow$ nuclear spin is flipped down among the (N - m) $\Uparrow$ nuclear spins and $Y_{\{m\}/l}(t)$ means that the l – th $\Downarrow$ nuclear spin among (m) $\Downarrow$ nuclear spins is flipped up.
From this construction, we may generalize the equations for the amplitudes when the m's are indexed in any order among the m nuclear $\Downarrow$ spins.

$$i\dot{Y}_{\{m\}}(t) = -B_{\{m\}}Y_{\{m\}} + \sum_{l\subset\{N-m\}}\left(\frac{A_l}{4}\right)X_{\{m+1\},l}(t) \quad (11)$$

$$i\dot{X}_{\{m+1\}}(t) = B_{\{m+1\}}X_{\{m+1\}}(t) + \sum_{l\subset\{m\}}\left(\frac{A_l}{4}\right)Y_{\{m\}/l}(t) \quad (12)$$

These simultaneous linear equations are solved by the Laplace transform method in time for arbitrary initial conditions. Note that $Y_{\{m\}}$ is a column matrix of size $\binom{N}{m}$ and $X_{\{m+1\}}$ is a column matrix of size $\binom{N}{m+1}$. The resulting equations are written compactly in terms of a partitioned matrix structure:

$$\left(\begin{array}{c|c} A_{\binom{N}{m}\otimes\binom{N}{m}} & C_{\binom{N}{m}\otimes\binom{N}{m+1}} \\ \hline D_{\binom{N}{m+1}\otimes\binom{N}{m}} & B_{\binom{N}{m+1}\otimes\binom{N}{m+1}} \end{array}\right)\left(\begin{array}{c} \bar{Y}_{\{m\}}(w) \\ \hline \bar{X}_{\{m+1\}}(w) \end{array}\right) = i\left(\begin{array}{c} Y_{\{m\}}(t=0) \\ \hline X_{\{m+1\}}(t=0) \end{array}\right)$$

(13)

Here we have introduced the following notations: $\bar{f}(w) = \int_0^\infty dt\, e^{-wt} f(t)$;

$A_{\binom{N}{m}\otimes\binom{N}{m}} = $ Diagonal matrix $(iw + B_{\{m\}})$

$B_{\binom{N}{m+1}\otimes\binom{N}{m+1}} = $ Diagonal matrix $(iw - B_{\{m+1\}})$

$C_{\binom{N}{m}\otimes\binom{N}{m+1}} = $ Rectangular matrix $\binom{N}{m} \otimes \binom{N}{m+1}$; and

$D_{\binom{N}{m+1}\otimes\binom{N}{m}} = $ Rectangular matrix $\binom{N}{m+1} \otimes \binom{N}{m}$ is the transposed matrix $C_{\binom{N}{m}\otimes\binom{N}{m+1}}$. This matrix equation has explicit solution:

$$\bar{Y}_{\{m\}}(w) = i(A - CB^{-1}D)^{-1}\left[Y_{\{m\}}(0) - CB^{-1}X_{\{m+1\}}(0)\right] \quad (14a)$$

$$\bar{X}_{\{m+1\}}(w) = i(B - DA^{-1}C)^{-1}\left[X_{\{m+1\}}(0) - DA^{-1}Y_{\{m\}}(0)\right] \quad (14b)$$

Equation (14b) is equivalently expressed in the form

$$C\bar{X}_{\{m+1\}}(w) = iY_{\{m\}}(0) - A\bar{Y}_{\{m\}}(w) \quad (14c)$$

This structure is transparent in exhibiting the same pole structure in both eqs. (14a) and (14b, c) which is useful in our later discussions concerning the time-dependence of the solutions. It is important to point out that the matrices *C* and *D* involve only the hyperfine interaction constants and so represent the effects of interactions in determining the amplitudes. Furthermore, the *i-th* diagonal element of the matrix $(A - CB^{-1}D)$ represents the self-energy contribution to the state whereas the zeros of the determinant of this matrix are the actual eigen-energies of the system. In [3], a continuum approximation is made even at the stage of eq.(13) yielding complicated integral equations. It is avoided here and the continuum approximation along with any other simplification are suggested to be made only at the end of the calculation.

The inversions of Laplace transforms yield the explicit time-dependencies of the solutions for any given initial specifications. In the next section we will illustrate the technique of this procedure in some simple choices of m which admit of generalization to more arbitrary m as well as N values.

## IV. Solutions for the amplitudes associated with a fixed $M_z(m)$ : an Illustrative example

We first consider in detail the special case of m=0 for which $M_z(m=0) = N-1$. This will already exhibit what is involved in finding the solutions of eqs.(14).

**m=0**:

Here eq.(13) takes a simple form because $Y_{\{0\}} = Y_0$ is a single number and $X_{\{1\}}$ is a column containing N elements indicated by a subscript $k = 1, 2, \ldots, N$. The various matrices are given by the following:

$$A_{\binom{N}{0} \otimes \binom{N}{0}} = \left(i\mathbf{w} + B_{\{0\}}\right), \quad B_{\binom{N}{1} \otimes \binom{N}{1}} = \text{Diagonal matrix}\left(i\mathbf{w} - B_{\{1\}}\right)$$

Here $B_{\{0\}} \equiv B_0 \equiv e + A$, and $B_{\{1\}} \equiv B_k \equiv B_0 - A_k$, $k = 1, 2, \cdots, N$.

$$C_{\binom{N}{m} \otimes \binom{N}{m+1}} = \text{Rectangular matrix}\binom{N}{0} \otimes \binom{N}{1} = \text{row } \frac{1}{4}(A_1 \ A_2 \cdots A_N);$$

$$D_{\binom{N}{1} \otimes \binom{N}{0}} = \text{Rectangular matrix}\binom{N}{1} \otimes \binom{N}{0} = \text{column } \frac{1}{4}\begin{pmatrix} A_1 \\ A_2 \\ \vdots \\ A_N \end{pmatrix}.$$

We then observe that

$$\left(A - CB^{-1}D\right) = (iw + B_0) - \frac{1}{16}\sum_{i=1}^{N}\left(A_i^2/(iw - B_0 + A_i)\right)$$

$$= \frac{\left\{(iw + B_0)\prod_{i=1}^{N}(iw - B_0 + A_i) - \frac{1}{16}\sum_{i=1}^{N}A_i^2 \prod_{j(\neq i)=1}^{N}(iw - B_0 + A_j)\right\}}{\prod_{i=1}^{N}(iw - B_0 + A_i)}$$

$$\equiv \prod_{l=1}^{N+1}(iw - \Omega_l) \Big/ \prod_{i=1}^{N}(iw - B_0 + A_i) \equiv \frac{D_{N+1}(w)}{d_N(w)} \quad (15)$$

where $\Omega_l$'s are the roots of the polynomial in the numerator and this combination is just a 1x1 matrix and we have introduced the notation

$$D_{N+1}(w) = \prod_{l=1}^{N+1}(iw - \Omega_l), \quad d_N(w) = \prod_{i=1}^{N}(iw - B_0 + A_i) \quad (16)$$

We thus obtain

$$\overline{Y}_0(w) = \frac{i}{D_{N+1}(w)}\left[d_N(w)Y_0(0) - \sum_{i=1}^{N}\frac{A_i}{4}X_i(0)d'_{N-1;A_i}(w)\right] \quad (17)$$

where we have introduced $d'_{N-1;A_i}(w) \equiv \prod_{l=1(\neq k_i)}^{N}(iw - B_0 + A_l)$. We also find

$$\overline{X}_j(w) = \frac{1}{(iw - B_0 + A_j)}\left[iX_j(0) - \frac{1}{4}A_j\overline{Y}_0(w)\right] \quad (j=1,2,\dots,N) \quad (18)$$

The inverse Laplace transforms give us the time-dependent solutions. An immediate point to note is that the initial state is obtained by the behavior of $\overline{Y}_0(w), \overline{X}_j(w)$ for large $w$ and both of these functions go as $w^{-1}$ as they should. The explicit evaluation of the Laplace transforms involve the roots $\Omega_l$ of the polynomial $D_{N+1}(w)$ in eq. (16). Note that these are exact solutions to the problem for m=0, $M_z(m=0) = N-1$. This structure is quite general. The resulting solutions display certain elegant symmetry which arise from the symmetric polynomials of the roots of the determinant of the matrix in eq. (13) or eq. (14a). Similar solutions can be obtained in all the cases

For any value of $m$ we have to solve for $\begin{pmatrix} \bar{Y}_{\{m\}}(w) \\ \rule{1cm}{0.4pt} \\ \bar{X}_{\{m+1\}}(w) \end{pmatrix}$ which obey simultaneous equations of order $N(M_z(m))$. The dimension of the matrix to be diagonalized is $N(M_z(m)) \times N(M_z(m))$. Formally, one could proceed by solving these equations and even invert the Laplace transforms to obtain the time-dependent solutions. But in practice, some numerical scheme is required to make the solution represented in graphical form. This is where the "pole approximation" suggests itself. As stated before, the matrices $C$ and $D$ represent the hyperfine interaction and may be dropped in computing the matrix inversion, which is the crux of the "pole approximation" [4]. We do not make any other approximation besides this, except in the final evaluation of physical quantities when the continuum approximation is invoked in obtaining their numerical estimations.

It may be of interest to note that if the number of nuclei is odd, then there is a state with $M_z = 0$ for $m = (N-1)/2$. For this, all the four matrices in eq.(13) are square matrices. On the other hand, if the number of nuclei is even, there are two states for which the nuclear spins add up to zero spin and $M_z = \pm 1/2$. The dimensions of the matrices $A, B$ are different and the matrices $C$ and $D$ will be rectangular.

In Appendix A, we work out the cases where N=3, 4 to demonstrate these features and how one should proceed to obtain such exact solutions. All these involve symmetric polynomials and identities concerning them and these explicit examples will exhibit this feature. These special cases also give clues to practical evaluations of the results of interest involving realistic computations.

## V. Summary and Concluding Remarks

We have presented here a formal solution to this much discussed problem of the electron dynamics in a quantum dot. Several issues concerning this problem in the light of the present work are discussed: (a) contrast with the master equation approach in [3]; (b) approximation scheme following from the exact analysis [4]; and (c) extension of the approach to include effects due to a second electron on another dot [5].

Coish and Loss [3] formulated the problem in terms of a master equation approach. They start with a factorized initial density matrix of the composite system of the electron and the nuclei at *t=0* and employ a particular projection operator method to develop a master equation. They use three types of initial density matrices for the nuclear systems for purposes of illustration. Recently there have been some suggestions for incorporating a non-factorized initial density matrix in developing non-Markovian mater equations [6]. In contrast, our method is for the pure state of the total system given arbitrary initial state of the composite state of the system. While the starting point of the mixed state is useful in considering for example, a system at finite temperature or an ensemble of states, the dependence of the results at subsequent evolution on the initial state is more complicated because of approximation schemes required in dealing with the master equation. The initial state of the system is experimentally prepared in feasible ways using various techniques (optical pumping, [7]) that can be more easily incorporated in the wave function method. Under these circumstances, the technique proposed here is more advantageous than the master equation procedure.

The approximation proposed in [4] is to employ the "pole approximation" after obtaining the exact solution outlined above. This takes advantage of the relative smallness of the individual hyperfine contributions in comparison with their sum, that sets the scale for the problem and the exact solution delineates these features neatly as discussed in the text, after eq.(14a, 14c). The use of continuum scheme to evaluate the sums that appear in the solution is postponed after obtaining the solution in the pole approximation. If the continuum scheme is used in setting up the equation as in [3], one obtains singularity structures that do not plague the calculations in [4].

The method employed here can be extended to include the effects of a second electron in a neighboring quantum dot. This problem has some new aspects such as quantum entanglement of the states of the two electrons, effects of degenerate states, etc. In [5], such a discussion is presented.

It may be worth pointing out that a general density matrix approach to this problem can be constructed by defining the total system density matrix in the form

$$\rho_{eN} = |\uparrow\rangle\langle\uparrow| \otimes A_N + |\uparrow\rangle\langle\downarrow| \otimes B_N + |\downarrow\rangle\langle\uparrow| \otimes B_N^+ + |\downarrow\rangle\langle\downarrow| C_N \quad (19)$$

where the coefficients are all matrices pertaining to the nuclei, $A_N, C_N$ are $2^N \times 2^N$ Hermitian matrices and $B_N$ is a $2^N \times 2^N$ matrix. By taking the trace over the nuclei, we obtain the marginal electron density matrix in the form (compare this with eqs.(4, 5) above)

$$\rho_e = \frac{1}{2}(I_2 + \vec{s}\cdot\vec{s}); \quad trA_N + trC_N = 1;$$
$$trA_N - trC_N = s_z; \quad trB_N = (s_x - i s_y)/2. \quad (20)$$

By taking the trace over the electron spins, we obtain the marginal nuclear density matrix

$$\rho_N = A_N + C_N \quad (21)$$

In place of the Schrodinger equation, one needs to solve the Liouville equation, $i\hbar_t \rho_{eN} = [H, \rho_{eN}]$, with given initial conditions where $H$ is the Hamiltonian in eq.(1). This gives us equations for the matrices in eq.(19) to be solved.

$$i\hbar_t A_N = \frac{1}{2}[g_z, A_N] + \frac{1}{2}(g_- B_N^+ - \hat{B}_N \hat{g}_+) \quad (21a)$$

$$i\hbar_t B_N = \frac{1}{2}[I_N \varepsilon + g_z, B_N]_+ + \frac{1}{2}(g_- C_N - A_N g_-), \quad (21b)$$

and

$$i\hbar\partial_t C_N = -\frac{1}{2}[g_z, C_N] + \frac{1}{2}(g_+ B_N - B_N^+ \hat{g}_-). \tag{21c}$$

In eqs.(21a,b) wee have used the notation $[A, B]_+ = AB + BA$. We have also introduced the notation $\vec{g} = \frac{1}{2}\sum_{k=1}^{N} A_k \vec{I}_k$. It should be understood that the operator $\vec{g}$ is an operator in the space of the composite of all the nuclear spins in the QD ($2^N$-dimensions in the present context). In this representation, $\vec{g} = \frac{1}{2}\sum_{k=1}^{N} A_k I \otimes \cdots \otimes \vec{I}_k \otimes I \cdots \otimes I$ where $I$ is the unit $2 \times 2$ matrix, where there are N factors in the above sum for each position of the nucleus. From eq.(20), all three matrices determine the electron spin dynamics. Notice from eq.(21) that the nuclear dynamics is contained in the matrices $A_N, C_N$, even though the Liouville equation would give rise to coupling of the three matrices. A development of this theory would encompass the mixed state situation of the system unlike the pure state development presented in detail here.

## APPENDIX A: EXPLICIT SOLUTIONS FOR N = 3, 4 CASES

**N = 3:**
Here we have 16 states, of which two of them are very simple and will not be discussed with $|\uparrow;\Uparrow\Uparrow\Uparrow\rangle\ (M_z = 4)\ \&\ |\downarrow;\Downarrow\Downarrow\Downarrow\rangle\ (M_z = -4)$. There are 4 states each with $M_z = \pm 2$ and 6 states with $M_z = 0$. We will discuss these in detail.

$M_z = 2$:
This is a special case discussed in Sec. 4 of the the text with N=3. The results are given below.

$$\bar{Y}_0(w) = \frac{i}{D_4(w)}\left[d_3(w)Y_0(0) - \sum_{i=1}^{3}\frac{A_i}{4} X_i(0)d'_{2;A_i}(w)\right] \tag{A1}$$

$$\bar{X}_j(w) = \frac{1}{(iw - B_0 + A_j)}\left[iX_j(0) - \frac{1}{4}A_j\bar{Y}_0(w)\right] \quad (j=1,2,3) \tag{A2}$$

$$D_4(w) \equiv \left\{(iw + B_0)d_3(w) - \frac{1}{16}\sum_{i=1}^{3} A_i^2\, d'_{2;A_i}(w)\right\} \tag{A3}$$

$$D_4(w) = \prod_{l=1}^{4}(iw - \Omega_l), \quad d_3(w) = \prod_{i=1}^{3}(iw - B_0 + A_i),$$

$$d'_{2;A_i}(w) \equiv \prod_{l=1(\neq k_i)}^{3}(iw - B_0 + A_l) \tag{A4}$$

We now invert the Laplace transforms to obtain explicit time dependence of the amplitudes. This is done here in detail to show how the symmetric polynomials play an important role in the solution. Introduce the symmetric polynomial

$$D_{4S} \equiv \prod_{i<j=1}^{4} (\Omega_i - \Omega_j) \tag{A5}$$

Then we obtain after picking up all the poles of $D_4(w)$ given in eq.(A4),

$$Y_0(t) = \frac{1}{D_{4S}} \sum_{l=1}^{4} (-1)^l e^{-it\Omega_l} \prod_{i<j(\neq l)}^{4} (\Omega_i - \Omega_j) [\tilde{d}(\Omega_l)] \tag{A6}$$

where $\tilde{d}(\Omega_l) \equiv \left[ d_3(\Omega_l) Y_0(0) - \sum_{j=1}^{4} \frac{A_j}{4} X_j(0) d'_{2,A_j}(\Omega_l) \right]$ (A7)

and

$$X_j(t) = e^{-it(B_0-A_j)} \left[ X_j(0) - \frac{A_j}{4} \overline{Y}_0(iw = (B_0 - A_j)) \right]$$

$$- \frac{A_j}{4} \frac{1}{D_{4S}} \sum_{l=1}^{4} (-1)^l \frac{e^{-it\Omega_l}}{\Omega_l - B_0 + A_j} \prod_{i<j(\neq l)}^{4} (\Omega_i - \Omega_j)[\tilde{d}(\Omega_l)]$$

(j=1,2,3)  (A8)

The various identities associated with symmetric polynomials play interesting roles in the symmetry of the solutions.

$M_z = 0$:

This has a different structure from the above. Also, the partitioned matrices in eq.(13) are all 3x3 square matrices. Here N=3 and m=1, and we index the nuclear states as follows:

$$|\Psi_{M_z=0}(t)\rangle = Y_1(t)|\downarrow;\Downarrow_1 \Uparrow_2 \Uparrow_3\rangle + Y_2(t)|\downarrow;\Uparrow_1 \Downarrow_2 \Uparrow_3\rangle + Y_3(t)|\downarrow;\Uparrow_1 \Uparrow_2 \Downarrow_3\rangle +$$
$$+ X_1(t)|\uparrow;\Uparrow_1 \Downarrow_2 \Downarrow_3\rangle + X_2(t)|\uparrow;\Downarrow_1 \Uparrow_2 \Downarrow_3\rangle + X_3(t)|\uparrow;\Downarrow_1 \Downarrow_2 \Uparrow_3\rangle$$
(A9)

Then the terms appearing in this case associated with the equation corresponding to eq.(13) has the form

$$A = \text{diagonal 3x3 matrix } (iw + B_0 - A_k); k = 1, 2, 3$$

$$B = \text{diagonal 3x3 matrix } (iw - B_0 + A_l + A_m); (l,m) = (2,3), (1,3), (1,2)$$
(A10a)

and

$$C = D = -\frac{1}{4}\begin{pmatrix} 0 & A_3 & A_2 \\ A_3 & 0 & A_1 \\ A_2 & A_1 & 0 \end{pmatrix} \quad (A.10b)$$

The column matrices representing the amplitudes are $\begin{pmatrix} \bar{Y}_1 \\ \bar{Y}_2 \\ \bar{Y}_3 \end{pmatrix}$, $\begin{pmatrix} \bar{X}_1 \\ \bar{X}_2 \\ \bar{X}_3 \end{pmatrix}$ and the resulting matrix equation leads to a 6$^{th}$ order polynomial determining the time-dependence of the solutions. For this problem, we need not resort to the partition matrix scheme as a direct inversion of this 6x6 matrix is possible. We may note that

$$[A - CB^{-1}D]\bar{Y}(w) = iY(0) - iCB^{-1}X(0) \quad (A11)$$

By examining the details of the matrix elements of $[A - CB^{-1}D]$ we note the following:
A typical diagonal element is of the form

$$(iw + B_0 - A_1) - \frac{1}{16}\left(\frac{A_3^2}{iw - B_0 + A_3 + A_1} + \frac{A_2^2}{iw - B_0 + A_2 + A_1}\right);$$

which is a self-energy contribution to the first term and a typical off-diagonal element has the form $-\frac{A_1 A_2}{16(iw - B_0 + A_1 + A_2)}$. For large $w$, $[A - CB^{-1}D]$ goes as $iw$. The typical term in RHS of eq. (A11) has the form $a_1(w) \equiv iY_1(0) - \frac{i}{4}\left\{\frac{A_3 X_2(0)}{iw - B_0 + A_3 + A_1} + \frac{A_2 X_3(0)}{iw - B_0 + A_2 + A_1}\right\}$. It is important to note that for large $w$ this goes as $a_1(w) \equiv iY_1(0) - \frac{i}{4(iw)}\{A_3 X_2(0) + A_2 X_3(0)\}$.

These observations ensure us that we recover the correct initial state at t=0.
The solution of eq.(14c) has a typical form

$$\bar{X}_1(w) = -\frac{2}{A_2 A_3}[A_3 b_3(w) + A_2 b_2(w) - A_1 b_1(w)],$$

where $b_i(w) = iY_i(0) - (iw + B_0 - A_i)\bar{Y}_i(w)$. For large $w$, terms cancel and we recover the correct initial state at t=0. This shows that there is no danger of large contributions arising from the hyperfine interactions appearing in the denominators.

**N = 4:**
There are 32 states in all in this case. The nuclei are labeled 1, 2, 3, 4 without loss of generality. We will not consider two of them, corresponding to $M_z = \pm 5$, for which both the electron and the nuclei all point in the same direction, either all up or all down. When the electron spin is down and all the nuclei are up, we have $M_z = +3$, and this is the case considered in Sec.IV. There are 5 states contributing to this case, $\{Y_0, X_k\}, k = 1, 2, 3, 4$. There is a corresponding case with $M_z = -3$, which is similar and we will not consider this here. When the electron spin is down and one of the nuclei is also down, we have 4 such states with $M_z = +1$ and this has 6 partners where the electron spin is up and two of the nuclei are up. This is an interesting case where the nuclear spins are compensated. This involves states to be labeled as follows:

$$\{Y_k; X_{lm}\}, k = 1, 2, 3, 4; (l, m) \text{ chosen from } \binom{4}{2} = 6 \text{ combinations of } (1, 2, 3, 4)$$

We will give the complete solutions explicitly in these two cases.

$M_z = 3$:

From Sec.4, we have

$$\overline{Y}_0(w) = \frac{i}{D_5(w)} \left[ d_4(w) Y_0(0) - \sum_{i=1}^{4} \frac{A_i}{4} X_i(0) d'_{3;A_i}(w) \right] \tag{A12}$$

$$\overline{X}_j(w) = \frac{1}{(iw - B_0 + A_j)} \left[ iX_j(0) - \frac{1}{4} A_j \overline{Y}_0(w) \right] \quad (j=1,2,3,4) \tag{A13}$$

$$D_5(w) \equiv \left\{ (iw + B_0) d_4(w) - \frac{1}{16} \sum_{i=1}^{4} A_i^2 d'_{3;A_i}(w) \right\} \tag{A14}$$

$$D_5(w) = \prod_{l=1}^{5} (iw - \Omega_l), \quad d_4(w) = \prod_{i=1}^{4} (iw - B_0 + A_i),$$

$$d'_{3;A_i}(w) \equiv \prod_{l=1(\neq k_i)}^{4} (iw - B_0 + A_l) \tag{A15}$$

$M_z = 1$:

The matrix structures here are different. This involves states to be labeled as follows:

$$\{Y_k; X_{lm}\}, k = 1, 2, 3, 4; (l, m) \text{ chosen from } \binom{4}{2} = 6 \text{ combinations of } (1, 2, 3, 4)$$

Explicitly

$$\left| \Psi_{M_z=1}(t) \right\rangle = \sum_{k=1}^{4} Y_k(t) \left| \downarrow; \{\Uparrow_k\} \right\rangle + \sum_{l<m=1}^{4} X_{lm}(t) \left| \uparrow; \{\Downarrow_l \Downarrow_m\} \right\rangle \tag{A16}$$

The matrix structure of A and B are 4x4 and 6x6 diagonal matrices respectively and C is a 4x6 matrix:

$$A = \text{diagonal 4x4 matrix } (i\mathbf{w} + B_0 - A_k); k = 1, 2, 3, 4$$

$$B = \text{diagonal 6x6 matrix } (i\mathbf{w} - B_0 + A_l + A_m); \quad \text{(A.17a)}$$

$$(l, m) = (1,2), (1,3), (1,4), (2,3), (2,4), (3,4)$$

$$C = -\frac{1}{4} \begin{pmatrix} A_2 & A_1 & 0 & 0 \\ A_3 & 0 & A_1 & 0 \\ A_4 & 0 & 0 & A_1 \\ 0 & A_3 & A_2 & 0 \\ 0 & A_4 & 0 & A_2 \\ 0 & 0 & A_4 & A_3 \end{pmatrix} \quad \text{(A17b)}$$

and $D = $ transpose of C.

The rest of the procedures are the same as before.

## ACKNOWLEDGEMENTS

Thanks are due to Dr. Tom Reinecke for focussing my attention to this problem. Dr. R. W. Rendell read this paper, drew my attention to ref.6., and made useful suggestions during the course of this research. This work is partially supported by the Office of Naval Research.